\documentclass[12pt]{article}
\usepackage{amssymb}
\usepackage{amsmath, amsthm}
\newcommand{\be}{\begin{equation}}
\newcommand{\ee}{\end{equation}}
\begin{document}
\def\theequation{\arabic{section}.\arabic{equation}}
\begin{titlepage}
\title{The stability of modified gravity models}
\author{Valerio Faraoni and Shahn Nadeau\\ \\
{\small \it Physics Department, Bishop's University}\\
{\small \it Lennoxville, Qu\`{e}bec, Canada J1M 1Z7}
}
\date{} \maketitle
\thispagestyle{empty}
\vspace*{1truecm}
\begin{abstract}
Conditions for the existence and stability of de Sitter space in 
modified gravity are derived by considering inhomogeneous 
perturbations in a gauge--invariant formalism. The stability 
condition coincides with the corresponding condition for 
stability with respect to homogeneous perturbations, while this 
is not the case in scalar--tensor gravity. The stability 
criterion is applied 
to various modified gravity models of the early and the 
present universe.
\end{abstract} \vspace*{1truecm}
\begin{center} {\bf PACS:}   98.80.-k, 04.90.+e, 04.50.+h
\end{center}
\begin{center}  {\bf Keywords:} modified gravity, de Sitter 
space, scalar--tensor gravity
\end{center}
\setcounter{page}{0}
\end{titlepage}

\def\theequation{\arabic{section}.\arabic{equation}}


\section{Introduction}
\setcounter{equation}{0}
 \setcounter{page}{1}

The 1998 discovery that the expansion of the universe is 
accelerated, obtained with the study of type Ia supernovae 
\cite{SN}, has prompted many theoretical models to explain this 
phenomenon. Most of these models can be classified in three 
classes: dark energy models, modified gravity models, and 
brane--world models. In the first class it is assumed that a form 
of {\em dark energy} or {\em quintessence} of unknown nature has 
come to dominate the dynamics of the universe at recent times 
(redshifts $z\leq 1$). These models are usually explored in the 
context of Einstein's theory of general relativity 
\cite{quintessence}, or possibly in its scalar--tensor 
generalizations ({\em extended quintessence}) \cite{ST}. Dark 
energy 
must necessarily have exotic properties in order 
to generate acceleration. In the spatially flat 
Friedmann--Lemaitre--Robertson--Walker (``FLRW'') line element 
describing our universe according to the recent cosmic microwave 
background experiments \cite{CMB}, and given by 
\be \label{1} 
ds^2=-dt^2 +a^2(t) \left( dx^2+dy^2+dz^2 \right) 
\ee
in comoving  coordinates $\left(t,x,y,z \right)$, the 
acceleration equation \cite{footnote1}
\be
\frac{\ddot{a}}{a}=\frac{\kappa}{6}\left( \rho+3P \right) 
\ee
holds, where $\kappa=8\pi G$ and $\rho $ and $P$ are the total 
energy density and pressure of the cosmic fluid, respectively. 
An overdot denotes differentiation with respect to the comoving 
time $t$. If dark energy were the only form of energy of 
the universe, 
acceleration $\ddot{a}>0$ would require an exotic negative 
pressure $P<-\rho/3 $; and if ordinary matter and dark matter 
contributing to the dynamics are taken into account, the 
pressure of dark energy must be even more negative to 
compensate.  As a matter of fact, dark energy is even more 
exotic: the best 
fit to the supernovae data favours an extreme form of dark 
energy called {\em phantom energy} or {\em superquintessence} 
with $P<-\rho$ --- or with an effective equation of state 
parameter  $w\equiv P/\rho <-1$. Furthermore, the effective 
equation of state 
should evolve with time \cite{w}. If confirmed, this fact would 
definitely rule out the cosmological constant $\Lambda$  as an 
explanation of the cosmic  acceleration because $w_{\Lambda}=-1$ 
is strictly constant (the 
cosmological constant model  is anyway disfavoured because of the 
cosmological constant problem \cite{Weinberg1} and of the cosmic 
coincidence problem \cite{Weinberg2} that accompany it). Most 
 models of dynamical dark energy  are based on a scalar field 
$\phi$ rolling 
in a potential $V(\phi)$, a  way to implement cosmic 
acceleration that is well known from inflationary theory in the  
early 
universe \cite{LiddleLyth}. However, a canonical scalar field 
minimally coupled to the Ricci curvature $R$ in 
Einstein gravity cannot explain an equation of state parameter 
$w<-1$, which is equivalent to {\em superacceleration} 
$\dot{H}>0$, where $H\equiv \dot{a}/a$ is the Hubble parameter. 
(This name distinguishes a regime in which the Hubble parameter 
increases from an ``ordinary'' acceleration regime in which 
$\ddot{a}=a\left( \dot{H}+H^2 \right)>0$ and $\dot{H}\leq 0$.) 
In fact, the energy density and pressure of such a scalar field 
are 
\be \label{3}
\rho_{\phi}=\frac{\dot{\phi}^2}{2} +V(\phi) \;, \;\;\;\;\;\;\;
P_{\phi}=\frac{\dot{\phi}^2}{2} - V(\phi)\;, \;\;\;\;\;\;\;
\ee
and the Einstein--Friedmann equation of general relativity
\be \label{4}
\dot{H}=-\frac{\kappa}{6} \left( P+\rho \right) 
=-\frac{\kappa}{2} \, \dot{\phi}^2 
\ee
yields $\dot{H}\leq 0$ for a universe dominated by such a scalar 
(the upper bound $H=$const. is attained by de Sitter space). 
In order to model an equation of state parameter $w<-1$ 
corresponding to $\dot{H}>0$, which is the situation favoured by 
the observational data, a phantom field with negative kinetic 
energy \cite{phantom} or a scalar field coupled non--minimally 
to gravity \cite{NMCquintessence} have been used. Both of these 
theories can be seen as special cases  of scalar--tensor gravity, 
described by the action
\be \label{5}
S=\int d^4x \sqrt{-g} \left[ \psi(\phi) R 
-\frac{\omega(\phi)}{2} \, 
g^{ab}\nabla_a \phi \nabla_b \phi -V(\phi) \right] \;,
\ee
where $\psi(\phi) $  and $\omega (\phi)$ are arbitrary coupling 
functions.  The 
exotic properties of dark energy or of its extreme form, phantom 
energy, have led some authors to a different approach and to the 
second class of models mentioned above. Instead of postulating 
an exotic form of dark energy of mysterious nature, these 
authors \cite{CCT,CDTT} consider the possibility that 
gravity deviates from Einstein gravity at large scales, and 
assume that the Einstein--Hilbert Lagrangian is modified by 
corrections that become important only at late times in the 
history of the universe, i.e., when the curvature becomes small. 
This class of theories, called {\em modified gravity}, is 
described by the gravitational action
\be \label{6}
S_g=\int d^4 x \sqrt{-g} \, f(R) \;,
\ee
where $f(R)$ is a non--linear function of $R$. The first model 
proposed had the form $f(R)=R-\mu^4/R$, in which the 
correction in $R^{-1}$ becomes important only at low curvatures 
$R\rightarrow 0$. The general form (\ref{6}) of the action also 
includes quantum gravity corrections to Einstein's theory 
originally introduced to improve renormalizability 
\cite{Stelle,Buchbinderetal} and used in inflationary models of 
the early universe \cite{Starobinsky80}. In addition to the 
desired phenomenological properties of modified gravity in 
cosmology, there is some motivation for these models from 
M--theory \cite{stringmotivations}. 

In both classes of models, depending on the arbitrary functions 
and parameters adopted, there are solutions describing universes 
that accelerate forever, other solutions in which the universe 
ends its existence in a finite time in the future in a Big Rip 
singularity \cite{BigRip} or encounters another type of ``sudden 
future  singularity'' 
\cite{sudden}. The fate of the universe depends on whether 
attractor solutions that are forever accelerating, or Big Rip 
attractors exist in the phase space, and on the size of their 
respective attraction basins. In many models of both dark energy 
and modified gravity there are de Sitter attractors accelerating 
forever. In this paper we focus on modified gravity and in 
sec.~3 we use scalar--tensor gravity for a comparison of 
properties. We determine whether de Sitter attractors exist in 
the phase space of modified gravity by deriving 
conditions for the existence and stability of these solutions. 
Throughout most of this paper we  
consider general non-linear actions of the form (\ref{6}) with 
$\partial^2  f/\partial R^2\neq 0$ and, in the final part of this 
paper, we apply our general results to specific scenarios and 
forms of $f(R)$ proposed in the literature.

While it is straightforward to study the stability of de 
Sitter space with respect to homogeneous perturbations, which 
only depend on time, it is physicallly more significant to 
assess stability with respect to more general inhomogeneous 
perturbations, which depend on both space and time. This goal is 
more ambitious because of the gauge--dependence problems 
associated with this  type of cosmological perturbations 
\cite{Bardeen} and one expects the stability condition with 
respect to inhomogeneous perturbations to be more restrictive 
than the corresponding condition for stability with respect to 
homogeneous perturbations. It comes therefore as  a surprise 
that these conditions coincide, as shown in sec.~3 and briefly 
reported in a previous communication \cite{RapidCommunication} 
--- this result can not be guessed or 
justified {\em a priori}. A similar analysis shows instead 
that in scalar--tensor 
theories the stability condition with respect to 
homogeneous perturbations is indeed more restrictive than the 
corresponding one for homogeneous perturbations  
\cite{RapidCommunication}. 

Certain modified gravity models are ruled out on the basis of 
instabilities that manifest on short timescales 
\cite{ChibaPLB03,DolgovKawasaki03,Chibagrqc0502070,
NunezSalganik,Wang05}. The 
stability condition 
derived here has the advantage of being applicable to any 
non-linear Lagrangian of the form $\sqrt{-g}\, f(R)$ and is 
useful in the study of the phase space and dynamics of modified 
gravity scenarios. Another motivation for our study is that, in 
order to be  viable, modified gravity models need to have the 
correct Newtonian  and post--Newtonian limit, and currently 
there is disagreement on whether certain models pass or not this 
test \cite{weakfield,Dick,CapozzielloTroisi,Sotiriou}. Because 
many  models do not admit a Minkowski solution around which to 
expand 
the weak--field metric, an expansion around the de Sitter 
background is used instead 
\cite{Dick,Chibagrqc0502070,NunezSalganik}. This is 
meaningful when a de Sitter solution exists and is stable.

The plan of this paper is the following: in sec.~2 we derive a  
stabiliy condition for de Sitter space with respect to 
inhomogeneous perturbations  by using a 
covariant and gauge--invariant formalism suitable for 
generalized gravity (including scalar--tensor and modified 
gravity, and possibly mixed models). In sec.~3 we derive the 
much simpler  stability condition with respect 
to homogeneous perturbations in modified gravity, and we 
interpret our results. In sec.~4 the stability condition derived 
for  modified gravity is applied to various scenarios widely 
discussed in the literature, while sec.~5 contains a discussion 
and the conclusions.


\section{Stability of de Sitter space with respect to 
inhomogeneous perturbations}
\setcounter{equation}{0} 

We can consider at once modified gravity and scalar--tensor 
theories, which we will use for a comparison with modified 
gravity, by studying the gravitational action
\be \label{7}
S = \int d^4 x \, \sqrt{-g} \left[ \frac{1}{2} \, f( \phi, R )  
-\frac{ \, \omega( \phi ) }{2}\,  g^{ab} \nabla_a \, \phi \,  
\nabla_b \,\phi -V( \phi) \right] \;.
\ee
This action contains also possible combinations of modified and 
scalar--tensor gravity if both $\partial f/\partial \phi$ and 
$\partial^2 f/\partial R^2 $ are non--vanishing: such mixed 
scenarios have received little attention in the literature so 
far \cite{mixedscenarios}. 
Scalar--tensor gravity is the special case in which $f$ is 
linear in $R$, i.e., $f \left( \phi, R \right)=\psi(\phi) R$, 
while modified gravity corresponds to setting $\phi=1$ and 
$\partial^2 f/\partial R^2\neq 0$. In the spatially flat FLRW 
metric (\ref{1}) the field equations assume the form
\begin{eqnarray} 
H^2 & =&  \frac{1}{3F} \left( \frac{\omega}{2} \, \dot{\phi}^2 
+\frac{RF}{2} -\frac{f}{2} +V 
-3H\dot{F}  \right)  \;, \label{8} \\
&& \nonumber \\
\dot{H} & = &  - \, \frac{1}{2F} \left(  \omega \dot{\phi}^2 + 
\ddot{F}  -H\dot{F} \right)  \;,
 \label{9} 
\end{eqnarray}

\be \label{10}
\ddot{\phi } +3 H \dot{\phi} +\frac{1}{2\omega} \left(
\frac{d\omega}{ d\phi} \,  \dot{\phi}^2 - \frac{\partial 
f}{\partial \phi} +2\, \frac{dV}{d\phi}  
\right) =0 \;,
\ee
where $ F \equiv \partial f / \partial R $. It is natural to use 
$\left( H,\phi \right)$ as dynamical variables and the 
equilibrium points of the dynamical system (\ref{8})--(\ref{10}) 
are de Sitter spaces with constant scalar field $\left( H_0, 
\phi_0 \right)$: they exist subject to the conditions
\be \label{11}
R_0 F_0 =2\left( f_0 -V_0 \right) \;,
\ee
\be \label{12}
f_0' = 2 V_0'  \;,
\ee
where $R_0=12H_0^2$, $f_0\equiv f\left( \phi_0, R_0 
\right) $, $F_0\equiv F\left( \phi_0, R_0 \right)$, 
$V_0=V(\phi_0)$, $V_0'= \left. 
\frac{dV}{d\phi}\right|_{\phi_0}$, and a prime denotes 
differentiation with respect to $\phi$. In modified gravity, 
there is only the condition (\ref{11}) for the existence of de 
Sitter solutions because there is only one arbitrary function 
$ f(R)$.

In order to describe inhomogeneous perturbations of the de 
Sitter fixed points $\left(H_0, \phi_0 \right)$ we use the 
covariant and gauge--invariant formalism of 
Bardeen--Ellis--Bruni--Hwang--Vishniac \cite{Bardeen,EllisBruni} 
in the  version given by 
Hwang and Hwang and Noh \cite{Hwang} for generalized gravity. 
The metric perturbations $A,B, H_L, $ and $H_T$ are defined by 
the relations
\begin{eqnarray} 
g_{00} & = & -a^2 \left( 1+2AY \right) \;, \label{13} \\
&& \nonumber \\
g_{0i} & = & -a^2 \, B \, Y_i  \;, \label{14} \\
&& \nonumber \\
g_{ij} & =& a^2 \left[ h_{ij}\left(  1+2H_L \right) +2H_T \, 
Y_{ij}  \right] \;,\label{15}
\end{eqnarray}
where the scalar harmonics $Y$ are the eigenfunctions of the 
eigenvalue problem $ \bar{\nabla_i}
\bar{\nabla^i} \, Y =-k^2 \, Y $, and where  $h_{ij} $ is the 
three--dimensional metric of the FLRW 
background, $ \bar{\nabla_i} 
$ is the covariant derivative associated with  $h_{ij}$, while 
$k$ is the  eigenvalue. The vector and tensor  harmonics $Y_i$ 
and $Y_{ij}$ 
satisfy the equations
\begin{eqnarray} 
Y_i & = &  -\frac{1}{k} \, \bar{\nabla_i} Y \;, \label{16}\\
&& \nonumber \\
Y_{ij} & =&   \frac{1}{k^2} \, \bar{\nabla_i}\bar{\nabla_j} Y 
+\frac{1}{3} \, Y \, h_{ij} \;.  \label{17}
\end{eqnarray}
 We need the  Bardeen gauge--invariant 
potentials \cite{Bardeen}
\begin{eqnarray} 
 \Phi_H & =&  H_L +\frac{H_T}{3} +\frac{ \dot{a} }{k} \left( 
 B-\frac{a}{k} \, \dot{H}_T \right) \;, \label{18}\\
&&\\
 \Phi_A & =&  A  +\frac{ \dot{a} }{k} \left( B-\frac{a}{k} \, 
\dot{H}_T \right)
 +\frac{a}{k} \left[ \dot{B} -\frac{1}{k} \left( a \dot{H}_T 
\right)\dot{}  \right] \;, \label{19}
\end{eqnarray}
and the  Ellis--Bruni variable  \cite{EllisBruni} 
\be \label{20}
 \Delta \phi = \delta \phi  +\frac{a}{k} \, \dot{\phi}  \left( 
B-\frac{a}{k} \, \dot{H}_T 
\right) \;,
\ee
while equations similar to eq.~(\ref{20}) define the 
gauge--invariant variables $\Delta f, \Delta F $, and 
$\Delta R$. The first order equations  satisfied 
by the gauge--invariant perturbations are  \cite{Hwang}
\begin{eqnarray}  \label{21}
&& \Delta \ddot{\phi} +  \left( 3H + \frac{\dot{\phi}}{ \omega} 
\, \frac{d\omega}{d\phi} \right) \Delta 
\dot{\phi} + \left[ \frac{k^2}{a^2}
 +  \frac{\dot{\phi}^2}{2} \frac{d}{d\phi} \left( 
\frac{1}{\omega} \frac{d\omega}{d\phi} \right) -\, 
\frac{d}{ d\phi} \left( \frac{1}{2\omega} \frac{\partial 
f}{\partial \phi} -\frac{1}{\omega} 
\frac{dV}{d \phi} \right) \right] \Delta \phi  \nonumber \\
&& \nonumber \\
&& =
  \dot{\phi}  \left(  \dot{\Phi}_A - 3\dot{\Phi}_H \right) 
 + \frac{\Phi_A}{\omega}  \left( \frac{\partial f}{\partial 
\phi} -2 \, \frac{dV}{d\phi} \right)  
 +\frac{1}{2\omega} \, \frac{\partial^2 f}{\partial \phi 
\partial  R}  \, \Delta R  \; ,
\end{eqnarray}

\begin{eqnarray}  \label{22}
&& \Delta \ddot{F} +3H \Delta  \dot{F} +\left( \frac{k^2}{a^2} - 
\frac{R}{3} \right) \Delta F 
+\frac{F}{3} \, \Delta R + \frac{2}{3} \, \omega \dot{\phi} 
\Delta \dot{\phi}  +\frac{1}{3} \left( \dot{\phi}^2 
\frac{d\omega}{d\phi}  + 2\frac{\partial f}{\partial \phi} 
-4 \, \frac{dV}{d\phi} \right) \Delta \phi \nonumber \\
&& \nonumber \\
&& = \dot{F}  \left(  \dot{\Phi}_A - 3\dot{\Phi}_H \right) 
+ \frac{2}{3}  \left( FR -2f +4V  \right)  \Phi_A \; ,
\end{eqnarray}

\be \label{23}
\ddot{H}_T +\left( 3H+  \frac{\dot{F}}{F} \right) \dot{H}_T 
+\frac{k^2}{a^2} \, H_T=0 \;,
\ee

\be  \label{24}
- \dot{\Phi}_H +\left( H +  \frac{\dot{F}}{2F} \right) \Phi_A = 
\frac{1}{2} \left(  \frac{  \Delta \dot{F} }{F} -H \frac{ \Delta F}{F} +  
\frac{ \omega}{F} \,  \dot{\phi} \, \Delta \phi \right)  \; ,
\ee

\begin{eqnarray}  \label{25}
& & \left( \frac{k}{a} \right)^2 \Phi_H +\frac{1}{2}
\left( \frac{ \omega}{F }  \dot{\phi}^2  + \frac{3}{2} 
\frac{\dot{F}^2}{F^2}  \right) \Phi_A =
\frac{1}{2} \left\{ \frac{3}{2}  \frac{ \dot{F} \Delta \dot{F} 
}{F^2} + \left( 3\dot{H} -  \frac{k^2}{a^2} -\frac{3H}{2} \frac{ 
\dot{F}}{F}  \right)  \frac{ \Delta F}{F} 
\right. \nonumber \\
&& \nonumber \\
& & \left. +\frac{\omega}{F}  \dot{\phi} \Delta \dot{\phi} + 
\frac{1}{2F} \left[ \dot{\phi}^2 
\frac{d\omega}{d\phi}  -\frac{ \partial f}{\partial \phi} 
+2\frac{dV}{d\phi} +6\omega \dot{\phi} 
\left( H +  \frac{ \dot{F} }{2F} \right) \right] \Delta \phi \right\} 
\; ,
\end{eqnarray}

\be   \label{26}
\Phi_A + \Phi_H =  - \frac{\Delta F }{F} \; ,
\ee

\begin{eqnarray}  
& & \ddot{\Phi}_H + H \dot{\Phi}_H + \left( H + \frac{ 
\dot{F}}{2F} \right) 
\left( 2\dot{\Phi}_H -\dot{\Phi}_A \right) 
+\frac{ 1 }{2F}  \left( f-2V -RF \right)  \Phi_A \nonumber \\ 
&& \nonumber \\
& & =  - \frac{1}{2} \left[ 
\frac{  \Delta \ddot{F}}{F} + 2H \, \frac{\Delta \dot{F}}{F} 
+ \left( P-\rho \right)  \frac{ \Delta F}{2F} + \frac{ 
\omega}{F} \, \dot{\phi} \, \Delta 
\dot{\phi} 
+ \frac{1}{2F} \left( \dot{\phi}^2  \, \frac{d\omega}{ d\phi}  
+\frac{\partial f}{\partial \phi 
}  -2 \, \frac{dV}{ d\phi } \right) \Delta \phi  \right]  \; 
, \nonumber \\
&&  
\end{eqnarray}
and
\be
\Delta R=6 \left[ \ddot{\Phi}_H +4H\dot{\Phi}_H 
+\frac{2}{3} 
\frac{k^2}{a^2} \Phi_H -H\dot{\Phi}_A 
-\left( 2\dot{H}+4H^2 -\frac{k^2}{3a^2} \right) \Phi_A \right] 
\;.    \label{28}
\ee
In the de Sitter background $\left( H_0, \phi_0 \right)$, these 
equations assume the considerably simpler form  
\be  \label{29}
\Delta \ddot{\phi} + 3H_0  \Delta \dot{\phi} 
+ \left[ \frac{k^2}{a^2}-\,  \frac{1}{2\omega_0} \left( f_0''   - 2 V_0'' \right) \right] 
\Delta \phi  =
 \frac{ f_{\phi R}}{2 \omega_0 } \,  \Delta R  \; ,
\ee

\be  \label{30}
\Delta \ddot{F} +3H_0 \, \Delta \dot{F} +\left( \frac{k^2}{a^2} - 4H_0^2 \right) \Delta F 
+\frac{F_0}{3} \, \Delta R =0 \;,
\ee

\be \label{31}
\ddot{H}_T +3H_0  \, \dot{H}_T +\frac{k^2}{a^2} \, H_T=0 \;,
\ee

\be \label{32}
-\dot{\Phi}_H+H_0 \Phi_A =\frac{1}{2} \left( \frac{\Delta \dot{F}}{F_0} -H_0 \, \frac{ \Delta 
F}{F_0} 
\right) \;,
\ee

\be  \label{33}
\Phi_H  = -  \frac{1}{2} \, \frac{ \Delta F}{F_0}   \; ,
\ee

\be   \label{34}
\Phi_A + \Phi_H =  - \frac{\Delta F }{F_0} \; ,
\ee

\be  \label{35}
\ddot{\Phi}_H + 3H_0 \dot{\Phi}_H  - H_0  \dot{\Phi}_A -3H_0^2 \Phi_A  
 =  - \frac{1}{2}\frac{  \Delta \ddot{F}}{F_0} - H_0 \, \frac{\Delta \dot{F}}{F_0} 
+ \frac{3H_0^2 }{2} \, \frac{\Delta F}{ F_0 }   \; ,
\ee
to first order, whereas
\be \label{36}
\Delta R=6  \left[ \ddot{\Phi}_H + 4H_0 \dot{\Phi}_H + 
\frac{2}{3} \frac{k^2}{a^2} \, \Phi_H 
-H_0 \dot{\Phi}_A  + \left( \frac{k^2}{3a^2} -4H_0^2 \right) 
\Phi_A \right] \;.
\ee
To first order and in the absence of ordinary 
matter \cite{footnote3}, vector perturbations 
do not have any effect. Expanding de Sitter spaces with  
$ H_0 > 0 $ are always stable, to first order, with respect to 
tensor  perturbations \cite{myPRD}, as can be seen from 
eq.~(\ref{23}). On  the other hand, contracting de Sitter 
spaces with $H_0<0$ 
are always unstable \cite{myPRD} and will not be considered 
further. There remain scalar perturbations, which we set out to 
examine.

In modified gravity theories with $f=f(R)$, $\phi \equiv 1$, and 
$f_{RR}\neq 0$, eqs.~(\ref{33}) and (\ref{34}) yield
\be \label{37}
\Phi_H=\Phi_A=-\, \frac{\Delta F}{2F_0} \;,
\ee
whereas eqs.~(\ref{36}) and (\ref{37}) yield
\be \label{38}
\Delta R=6\left[ \ddot{\Phi}_H+3H_0\dot{\Phi}_H +\left( 
\frac{k^2}{a^2}-4H_0^2 \right)\Phi_H \right] 
\ee
and $a=a_0 \, \mbox{e}^{H_o t}$ is the scale factor of the 
unperturbed de Sitter space, with $a_0$ a constant. In the de 
Sitter background the gauge--invariant variables reduce, to 
first order, to
\be \label{39bis}
\Delta \phi = \delta \phi \;, \;\;\;\; \Delta R=\delta R\; , 
\;\;\; \Delta F=\delta F \;, \;\;\;\; 
\Delta f =\delta f \;.
\ee
Since the function $F$ depends only on $R$ one has
\be \label{39}
\frac{\Delta F}{F_0}=\frac{ f_{RR}}{F_0}\, \Delta R
\ee
where 
\be
f_{R R} \equiv \left.  \frac{ \partial^2 f}{ \partial R^2} 
\right|_{ R_0 } \;,
\ee
and therefore
\be
\Delta R=-\, \frac{2F_0}{f_{RR}}\, \Phi_H \;.
\ee
The perturbations $\Phi_H=\Phi_A$ then evolve according to 
eq.~(\ref{35}), which becomes 
\be \label{41}
\ddot{\Phi}_H+3H_0\dot{\Phi}_H+\left( 
\frac{k^2}{a^2}-4H_0^2+\frac{F_0}{3f_{RR}}\right)\Phi_H=0 \;,
\ee
where $ a=a_0 \, \mbox{e}^{H_0 t}$, with $a_0$ a constant. At 
late 
times the term $k^2/a^2$ can be neglected and stability 
is achieved if the coefficient of $\Phi_H$ in the last term of 
the left hand side of eq.~(\ref{41}) is positive or zero: upon 
use of the value of the Hubble parameter given by 
eq.~(\ref{11}) for the unperturbed de Sitter space, this 
condition reduces to 
\be \label{42}
\frac{F_0^2-2f_0f_{RR}}{F_0f_{RR}} \geq 0 \;.
\ee
This inequality was presented in a previous communication 
without details of the derivation \cite{RapidCommunication}. We 
now  comment on the physical meaning of the approximation 
leading to 
(\ref{42}). Information on the spatial dependence of the 
inhomogeneous scalar perturbations are contained in the 
eigenvector $k$ of the spherical harmonics, and the fact that 
the only term containing $k$ in eq.~(\ref{41}) becomes 
negligible as time progresses in a de Sitter background implies 
that the spatial dependence effectively disappears from the 
analysis. One may be tempted to conclude that the stability 
condition (\ref{42}) with respect to inhomogeneous 
perturbations could be obtained in a much quicker way by 
considering the simpler homogeneous perturbations: this would 
be incorrect as one would not be able to guess {\em a priori}, 
in 
a homogeneous perturbation analysis,  
the structure of eq.~(\ref{41}) and the fact that the spatial 
dependence disappears. Furthermore, in the parallel case of 
scalar--tensor gravity, the stability condition with respect to 
homogeneous perturbations differs from the corresponding one for 
inhomogeneous perturbations, and one would expect the same to 
happen for modified gravity. This is the subject of the next 
section.

\section{Homogeneous perturbations in modified gravity and in 
scalar--tensor theories}
\setcounter{equation}{0}

We now derive the stability condition of de Sitter space 
with respect to {\em homogeneous} perturbations
in modified gravity, in order to compare it with the result 
(\ref{42}) of the previous section. The field equations reduce 
to
\begin{eqnarray}
H^2 & =& \frac{1}{3F} \left( \frac{RF-f}{2}-3H\dot{F} \right)\;, 
\label{43} \\
&& \nonumber \\
\dot{H} & = & -\, \frac{1}{2F} \left( \ddot{F}-H\dot{F} \right) 
\;. \label{44}
\end{eqnarray}
By assuming that $H(t)=H_0+\delta H(t) $ and using the first 
order expansions 
\begin{eqnarray}
&& R=R_0+\delta R \;, \;\;\;\;\;\;\delta R=6\left( 
\delta\dot{H} +4H_0\delta H  \right) \;, \nonumber \\
&& F=F_0+f_{RR}\delta R\;, \;\;\;\;\;\;\;
f=f_0+F_0\delta R  \;,\label{45}
\end{eqnarray}
and eq.~(\ref{11}), one obtains the evolution equation for the 
homogeneous perturbation $\delta H$
\be \label{46}
\delta \ddot{H}+\left( 4H_0-\frac{f_0}{6H_0F_0} \right) \delta 
\dot{H} 
+\frac{1}{3}\left( \frac{F_0}{f_{RR}}-\frac{2f_0}{F_0} \right) 
\delta H=0 \;.
\ee
The ansatz $\delta H=\epsilon \, \mbox{e}^{st} $ yields an  
algebraic equation for $s$ with roots
\be\label{47}
s_{\pm}=\frac{1}{2} \left[ -\frac{f_0}{2H_0F_0}\pm \sqrt{ \left( 
\frac{f_0}{2H_0F_0} \right)^2 -\frac{4}{3}\left( 
\frac{F_0}{f_{RR}}-\frac{2f_0}{F_0} \right)} \, \right] \;.
\ee
Assuming $f_0 >0$ and $H_0>0$, if also $F_0>0$, then 
$-f_0/\left( 2H_0F_0 \right)<0$ and there is stability if and 
only if
\be \label{48}
\frac{F_0}{f_{RR}}-\frac{2f_0}{F_0} \geq 0 \;,
\ee
which is equivalent to the condition (\ref{42}). If (\ref{48}) 
is not satisfied, the root $s_{+}$ is real and positive, 
corresponding to an unstable mode growing exponentially in time. 

The case $F_0<0$ does not correspond to a de Sitter solution 
when $f_0>0$ because eq.~(\ref{11}), which reduces to $R_0F_0 = 
2f_0$ cannot be satisfied in this case.

Why the stability condition with respect to homogeneous 
perturbations coincides with the corresponding stability 
condition with respect to inhomogeneous perturbations? A naive 
answer would be that the spatial dependence of the inhomogeneous 
perturbations can safely be eliminated in the analysis of 
eq.~(\ref{41}) or, in other words, inhomogeneities are 
redshifted away, and the results must coincide. This would be 
intuitive: even initial anisotropies are known to be smoothed 
out by de Sitter--like expansion \cite{Waldtheorem} in general 
relativity and in some scalar--tensor cosmologies \cite{Jorge}, 
but this is  not the correct explanation: in fact, if it were 
true, it should hold  also in the case of scalar--tensor gravity 
in which $f\left(  \phi, R \right)=\psi \left(\phi \right)R$, 
but this is not the  case as  we are going to show. The 
stability condition with respect to 
inhomogeneous perturbations in scalar--tensor gravity has been 
derived in Ref.~\cite{myPRD} by analysing the equation for the 
gauge--independent Bardeen potentials and for the Ellis--Bruni 
variable $\Delta \phi$,
\be \label{44bis}
\Delta \ddot{\phi}+3H_0\Delta \dot{\phi}+\left[ 
\frac{k^2}{a^2}-\frac{ \frac{f_0''}{2}-V_0''+\frac{6f_{\phi 
R}^2H_0^2}{F_0}}{\omega_0 \left( 1+ \frac{3f_{\phi 
R}^2}{2\omega_0 F_0}\right)}\right] \Delta \phi= 0 \;,
\ee
where
\be
 f_{\phi R} \equiv \left.  \frac{ \partial^2 f}{ \partial \phi 
\partial R} 
\right|_{ \left( \phi_0, R_0 \right)} \;, \;\;\;\;\;\;
f_0'' \equiv \left. \frac{\partial^2 f}{\partial \phi^2} 
\right|_{\left( \phi_0, R_0 \right)} \;.
\ee
This equation is  obtained from eqs.~(\ref{29})--(\ref{36}) if
$ 1+3f_{\phi R}^2/\left( 2\omega_0 F_0 \right) \neq 0$. The 
stability condition that ensues is \cite{myPRD}
\be \label{45bis}
 \frac{ \frac{f_0''}{2}-V_0''+\frac{6f_{\phi 
R}^2H_0^2}{F_0}}{\omega_0 \left( 1+ \frac{3f_{\phi 
R}^2}{2\omega_0 F_0}\right)}\leq 0 \;.
\ee
For the sake of comparison, let us derive the corresponding 
stability condition with respect to homogeneous perturbations in 
scalar--tensor gravity. Assuming that $ f\left( 
\phi, R \right)=\psi( \phi)R$, the homogeneous perturbations 
\be \label{46bis}
H(t)=H_0+\delta H(t) \;, \;\;\;\;\; \phi(t)=\phi_0+\delta 
\phi(t) 
\ee
satisfy the first order evolutions equations 
\begin{eqnarray} 
&& \delta \dot{H}=-\, \frac{1}{2\psi_0}\left( \psi_0'\delta 
\ddot{\phi}-H_0\psi_0'\delta \dot{\phi} \right) \;,\label{47bis} 
\\
&& \nonumber\\
&& 
\delta \ddot{\phi}+3H_0\delta 
\dot{\phi}+\frac{1}{2\omega_0}\left( 
2V_0''-f_0''\right)\delta \phi=0 \;. \label{48bis}
\end{eqnarray}
By contrast, in general relativity with a minimally coupled 
scalar field the perturbations have no effect to first order. 
This is due to the non--canonical form of the effective 
energy--momentum tensor of the scalar field appearing in the 
left hand side of the field equations of scalar--tensor gravity 
when these are written in the form $G_{ab}=8\pi 
T_{ab}^{(eff)}\left[ \phi \right] $ (see 
Refs.~\cite{mybook,BellucciFaraoni} for a discussion). The 
stability condition with respect to homogeneous perturbations 
can be read off of eq.~(\ref{48bis}),
\be \label{49}
\frac{\frac{f_0''}{2}-V_0''}{\omega_0} \leq 0 \;.
\ee
The stability condition (\ref{45bis}) with respect to 
inhomogeneous perturbations is more restrictive than (\ref{49}) 
and, in spite of having neglected a term $k^2/a^2=k^2 \, 
\mbox{e}^{-2H_0t}/a_0^2$ in eq.~(\ref{44bis}), the final 
stability  condition (\ref{45bis}) retains a memory of the 
spatial dependence 
of the inhomogeneous perturbations, which is instead lost in the 
homogeneous perturbation analysis leading to (\ref{49}). {\em A 
priori}, one should expect a similar situation for modified 
gravity, and the fact that the two stability conditions 
coincide for these theories appears to be  coincidental.

For the particular class of scalar--tensor theories described by 
the  action
\be\label{50}
S=\int d^4x \sqrt{-g} \left[ \phi R -\frac{\omega(\phi)}{\phi}\, 
g^{ab}\nabla_a\phi\nabla_b\phi -V(\phi) \right] 
\ee
and containing a single arbitrary coupling function 
$\omega(\phi)$, 
the stability conditions (\ref{45bis}) and (\ref{49}) coincide 
\cite{MikeJensen}. However, these two conditions fail to 
coincide in the general 
scalar--tensor theory (\ref{5}) because, in the right hand side 
of eq.~(\ref{29}), curvature perturbations $\Delta R$ act as 
sources for the perturbations $\Delta \phi$ (which, in the de 
Sitter background, coincide with $\delta \phi$). On the 
contrary, such a term is absent in the homogeneous perturbation 
analysis of the Klein--Gordon equation (\ref{10}), and an 
analogous term does not appear in eq.~(\ref{41}) for the 
gauge--independent Bardeen potential $\Phi_H$ in modified 
gravity because there is no scalar field in this case and 
$f_{\phi R}=0$ ---  eq.~(\ref{29}) then becomes homogeneous.

\section{Application to specific modified gravity scenarios}
\setcounter{equation}{0}

We now proceed to apply the stability condition 
(\ref{42}) to certain specific modified  gravity scenarios that 
have been proposed in the literature, for 
which a de Sitter space is relevant.

\subsection*{$\bullet$~~~ $ f(R)=R-\frac{\mu^4}{R}$}

This theory 
\cite{CCT,CDTT,ChibaPLB03,DolgovKawasaki03,stringmotivations,
MengWang}, with 
the mass scale $\mu_0\simeq H_0\simeq 10^{-33}$~eV, was  the 
first candidate proposed to explain the cosmic acceleration, 
and it is known to be subject to an instability that develops on 
a time scale of order $10^{-26}$~s 
\cite{DolgovKawasaki03}. For this theory, the stability condition 
(\ref{42}) reduces to
\be
1+\frac{6\mu^4}{R_0^2}-\frac{3\mu^8}{R_0^4} \leq 0 
\ee
and it is clear that by taking $\mu_0\approx H_0$ with 
$R_0=12H_0^2$, the stability 
condition is impossible to satisfy. More precisely, the 
condition for the existence of de Sitter solutions is 
$R_0=\sqrt{3}\, \mu^2 $ (see also Ref.~\cite{CDTT}) and the 
stability condition (\ref{42}) reduces to $8/3\leq 0$, which 
obviously cannot be satisfied. We stress that this instability 
of de Sitter space arises in the  gravitational  sector of the 
theory, while the instability discovered in 
Ref.~\cite{DolgovKawasaki03} arises in the matter sector
and would disappear in vacuum. The instability of de Sitter 
space can be stabilized by adding a term of the form $\epsilon 
R^2$ with $0< \epsilon < \mu^4$ to the Lagrangian density, as 
shown in the following.

\subsection*{$\bullet$~~~ $ f(R)=R-\frac{\mu^4}{R} +aR^2 $}

In this theory \cite{NojiriOdintsovPRD03} the condition for the 
existence of de Sitter solutions (\ref{11}) becomes 
\cite{NojiriOdintsovPRD03,Cognolaetal}
\be \label{52}
R_0=\sqrt{3} \, \mu^2 \;,
\ee
independent of the parameter of the quadratic correction --- 
therefore this condition holds true also for $a=0$ as seen in 
the 
previous case. For general values of the parameter $a$, upon 
use of eq.~(\ref{52}), the 
stability condition (\ref{42}) reduces to 
\be
\frac{1}{3\sqrt{3}\, a\mu^2-1}\geq 0 
\ee
and therefore de Sitter space is stable if 
$ a> \left( 3\sqrt{3}\, \mu^2 \right)^{-1} $ and unstable if 
$ a<  \left( 3\sqrt{3}\, \mu^2 \right)^{-1} $ (in particular 
for $a=0$, which is the previous case). Therefore, adding a 
quadratic correction with $ a < \left( 3\sqrt{3}\, \mu^2 
\right)^{-1}  $ (and in particular with a negative $a$, which 
reinforces the effect of the term $-\mu^4/R$) leads to 
instability. Now, if $\mu\sim H_0\sim 10^{-33}$~eV, the 
parameter $a$ must be larger than $\sim 10^{65} \, \left( 
\mbox{eV} \right)^{-2}$ for stability, which appears to be huge 
in natural units --- stability is achieved at the price of 
fine--tuning the parameters.

\subsection*{$\bullet$~~~ $ f(R)=R^n $}

This modified gravity theory has been pursued in the literature 
\cite{CCT,CarloniDunsbyCT}, especially for $n=-1$ and for 
$n=3/2$, in which case it is  is 
conformally equivalent to Liouville field theory
\cite{CCT}. The model has been used to explain the cosmic 
acceleration and it is also 
interesting because ordinary inflation with a minimally coupled 
scalar field and an exponential potential (power--law inflation) 
can be rewritten as a theory $f(R)=R^n$ 
\cite{NojiriOdintsovPRD03}. For generic values of $n\neq 
0,1/2,1$, the theory yields power--law inflation 
$ a\propto 
t^{\alpha}$ with 
\be
\alpha=\frac{-2n^2+3n-1}{n-2} \;.
\ee
This theory does not admit de Sitter solutions if $n\neq 2$ 
\cite{BarrowOttewill,myPRD} unless a cosmological constant,  
corresponding to a term with $n=0$, is added to the action 
\cite{BarrowOttewill}. However, Minkowski space (the trivial de 
Sitter space) is a solution without cosmological constant for 
any positive value  of $n$ \cite{footnote2}. In fact, the 
condition (\ref{11}) for the 
existence of de Sitter solutions yields $nR_0=2R_0^n$, which is 
only satisfied for $n=2$ or $R_0=0$. The stability condition 
(\ref{42}) yields
\be
R_0 \, \frac{2-n}{n\left( n-1 \right)} \geq 0 
\ee
and is satisfied for $1< n \leq 2 $ and for $n<0 $; it is 
identically satisfied for $n=2$, without imposing constraints on 
the Hubble parameter $ H_0 $ of de Sitter space. The Minkowski 
spaces $R_0=0$ are stable for any $n>0$.

\subsection*{$\bullet$~~~ $ f(R)=R +\epsilon R^2 $}

Quadratic corrections to the Einstein--Hilbert Lagrangian 
density are motivated by renormalizability 
\cite{Stelle,Buchbinderetal} and 
higher order corrections are unavoidable near the Planck scale 
\cite{Vilkovisky}. This theory was used in one of the earliest 
inflationary scenarios \cite{Starobinsky80}, not requiring an 
inflaton field. The constant has dimensions $\epsilon \sim 
M^{-2}$,  where $M\sim 10^{12} $~GeV. The condition for the 
existence of de Sitter solutions allows only the 
trivial Minkowski space 
$H_0=0$. However, there are non--trivial de Sitter solutions if 
a cosmological constant is added to $f(R)$ \cite{myPRD}. The 
stability condition (\ref{42}) yields 
\be
\frac{1}{\epsilon \left( 1+2\epsilon R_0 \right)} \geq 0
\ee
which, for Minkowski space, gives stability if $\epsilon >0$ and 
instability for $\epsilon<0$. The case $\epsilon=0$ 
corresponding to Einstein's theory must be studied separately 
and it is concluded that Minkowski space is stable in this case 
\cite{myPRD}.

\subsection*{$\bullet$~~~ $ f(R)=R +\epsilon R^2 -2\Lambda $}

By adding a cosmological constant to the previous theory, de 
Sitter solutions become possible and are given by $ 
R_0 = 4 \Lambda$,  or $H_0=\sqrt{\Lambda/3} $ as in general 
relativity, because the 
condition on $H_0$ does not depend on the parameter $\epsilon$ 
and coincides with the corresponding condition for $\epsilon=0$. 
The stability condition (\ref{42}) reduces to $\epsilon>0$: a 
positive quadratic correction acts in the same direction as the 
term $R$ in the Lagrangian, whereas a negative quadratic 
correction with $\left| \epsilon \right|$ arbitrarily small 
makes de Sitter space unstable.

\subsection*{$\bullet$~~~ $ f(R)=R + \epsilon R^n $}

This theory 
\cite{CDTT,CFDETT}, 
with $ \epsilon \sim M^{2\left( 1-n \right)}$, where 
$M$ is a mass scale, comprises quantum 
gravity--motivated corrections to the 
Einstein--Hilbert Lagrangian for $n>0$, as well as the theory 
$ f(R)=R-\mu^4/R $ already discussed, or similar theories, if 
$n<0$. The condition for the existence of de Sitter space is 
either $ R_0=0 $ (Minkowski space) or 
\be \label{Delta}
R_0^{n-1}=\left(12H_0^2 \right)^{n-1}=\frac{1}{\epsilon \left( 
n-2 \right)} 
\ee
for $n\neq 2$ (the case $ n =2 $ has already been considered). 
In 
the case of a non--trivial de Sitter space $H_0\neq 0$ the 
stability condition (\ref{42}) yields, using eq.~(\ref{Delta}),
\be
\frac{-n^2+2n-2}{\epsilon n \left( n-1 \right)}\geq 0
\ee
or $\epsilon \, n<0$, and therefore de Sitter space is stable if 
$\epsilon n<0 $ and unstable otherwise. In particular, it is 
stable if $ \epsilon >0 $ and $ n<0 $ (or if $\epsilon 
<0 $ and $ n> 0 $), which comprises the 
case $f(R)=R-\mu^4/R$ (as $n=2$ is excluded from this analysis, 
these results do not contradict the previous statements on the 
stability of Minkowski space when $n=2$). Therefore, any theory 
of the form $f(R)=R-\mu^{2\left( 1-n \right)} /R^m $ with $m>0$ 
exhibits the same instability in the gravitational sector as the 
the model 
$f(R)=R-\mu^4/R$ and, likely, the same 
instability reported in Ref.~\cite{DolgovKawasaki03} for the 
matter sector.

\subsection*{$\bullet$~~~ $ f(R)= a \ln \left( \frac{R}{b} 
\right) $}

This theory \cite{CapozzielloTroisi,NOGRG,MengWangPLB04}, which 
does not admit 
a 
Minkowski space or other 
solutions with vanishing Ricci curvature, admits a de Sitter 
solution only if $ R_0=12H_0^2=b\, \mbox{e}^{b/2} $, where $b$ 
is  a positive parameter. The stability condition (\ref{42}) 
reduces to
\be
b \left[ b+2\ln \left( \frac{R_0}{b} \right) \right] \leq 0
\ee
which, for $ b > 0 $, is equivalent to $\mbox{e}^b \leq 0$ and 
obviously is never satisfied: the de Sitter space $H_0=\sqrt{ 
b\, \mbox{e}^{b/2}/12} $ is unstable.


\section{Discussion and conclusions}
\setcounter{equation}{0} 

The general stability condition (\ref{42}) of de Sitter space in 
modified gravity derived in sec.~2 allows one to quickly assess 
the stability of de Sitter space in specific scenarios. Of 
course, when the number of terms in the Lagrangian density 
grows, so does the volume of parameter space to be searched and  
this analysis becomes cumbersome --- it would be  greatly helped 
by an  estimate of the range of values of the parameters 
involved. In the future we plan to extend  to 
power--law solutions the study carried out here for de Sitter 
spaces.

The comparison of the modified gravity results with the 
analogous results in scalar--tensor gravity shows that, although 
there exists a dynamical equivalence between modified gravity 
and scalar--tensor gravity \cite{TeyssandierTourrenc}, this 
should not be taken too literally. The stability conditions 
with respect to homogeneous and inhomogeneous perturbations 
coincide in modified gravity but not in scalar--tensor gravity, 
due to the different detailed structure of the equations 
satisfied by the perturbations.

The scope of the stability analysis can perhaps be extended to 
more general theories of the form $f\left( R, R^2, R_{ab}R^{ab}, 
R_{abcd}R^{abcd} \right)$ containing string--motivated 
corerctions (see, e.g., 
Refs.~\cite{DobadoMaroto,Brandenbergeretal,Samietal05}); 
there are however doubts on certain choices of the string 
corrections to the Einstein--Hilbert action, due to ghosts or 
light, long--ranged, gravitational scalars that potentially 
violate Solar System bounds 
\cite{NunezSalganik,Chibagrqc0502070}.

Finally, it should be stressed that, while we have analyzed 
linear stability with respect to inhomogeneous perturbations 
described by gauge--invariant variables, other definitions of 
stability can be considered: for example, stability with respect 
to black hole nucleation \cite{Cognolaetal} or quantum 
fluctuations \cite{DolgovPelliccia}. Sometimes these different 
definitions yield results that are qualitatively similar to our 
stability condition (\ref{42}) (e.g., \cite{Cognolaetal}). 
Another possibility is to search for a 
positive--definite energy functional \cite{Orfeu}, although one 
should probably look for energies bounded from below 
rather than  positive energies \cite{negenergy}; and so on. The 
fact that these stability criteria are inequivalent is not 
surprising since  the physical processes considered are quite  
different and, even from the mathematical point of view, 
several inequivalent definitions of stability exist for 
dynamical systems \cite{Glendinning}.

\section*{Acknowledgments}

This work was supported by a grant from the Senate Research 
Committee of Bishop's University. V.F. is also supported by the 
Natural Sciences and Engineering Research Council of Canada 
(NSERC).


\end{document}